# X-ray Constraints on the Intrinsic Shapes and Baryon Fractions of Five Abell Clusters

David A. Buote[1] and Claude R. Canizares[2]

Department of Physics and Center for Space Research 37-241,

Massachusetts Institute of Technology

77 Massachusetts Avenue, Cambridge, MA 02139

astro-ph/9504049 2 Aug 95

## ABSTRACT

We analyzed ROSAT PSPC images of the bright, nearby ($z < 0.1$) galaxy clusters A401, A1656 (Coma), A2029, A2199, and A2256 to constrain their intrinsic shapes and baryon fractions; the intrinsic shapes of these clusters were analyzed previously by us using *Einstein* data (Buote & Canizares 1992; Buote 1992). Following Buote & Tsai we probed the aggregate structure of the clusters on scales $\sim 1.5 h_{80}^{-1}$ Mpc to reduce effects of possible substructure on smaller scales ($\lesssim$ a few hundred kpc). The ellipticities of the X-ray isophotes are typically $\epsilon \approx 0.15 - 0.25$ and display negative radial gradients highly significant at the 95% confidence limit for all the clusters except Coma. By assuming hydrostatic equilibrium and a variety of mass models we obtain ellipticities $\epsilon_{mass} \approx 0.40 - 0.55$ for isothermal models of the total gravitating matter of the clusters; the $\epsilon_{mass}$ constraints change by $< 10\%$ upon consideration of the small temperature gradients shown by ASCA to be typical for rich clusters; the observed X-ray ellipticity gradients require that mass models with constant ellipticity have steep density profiles, $\rho \propto r^{-4}$. Estimates of the gas masses are highly insensitive to the ellipticities of the X-ray isophotes. The clusters in our sample have increasing fractions of gas mass to total mass with radius and have $M_{gas}/M_{tot} = (4\% - 11\%) h_{80}^{-3/2}$ within a radius $1.5 h_{80}^{-1}$ Mpc, in excellent agreement with the results of White & Fabian and the Baryon Catastrophe proposed by White et al. (1993). Finally, the ellipticities of the dark matter distributions are essentially identical to $\epsilon_{mass}$ and are consistent with the shapes of dark halos predicted by N-body simulations and the shapes of the galaxy isopleths in the clusters in contrast to our previous conclusions using *Einstein* data.

---

[1]dbuote@space.mit.edu

[2]crc@space.mit.edu

## 1. Introduction

The structures of galaxy clusters probe cosmological theories in a variety of ways. The influence of early tidal distortions in the formation of clusters (Binney & Silk 1979), the degree of dissipation of the dark matter (e.g., Strimple & Binney 1979; Aarseth & Binney 1978), and the density of the universe (Eisenstein & Loeb 1995) all affect the intrinsic shapes of clusters. Morphological differences between the X-ray–emitting gas and the underlying mass in clusters may reflect the nature of the dark matter itself (Kaiser 1991). Finally, the large baryon fractions observed for Coma (Briel, Henry, & Böhringer 1991) and other clusters (White & Fabian 1995) may have serious repercussions for many cherished notions of the standard cosmology (White et al. 1993).

We (Buote & Canizares 1992 – hereafter BC; Buote 1992) have previously analyzed the shapes of a small sample of bright, low-redshift ($z < 0.1$) Abell clusters having no obvious subclustering (with the possible exception of Coma) using the *Einstein* Imaging Proportional Counter (IPC; Giacconi et al. 1979) and concluded that the ellipticities of the gravitating mass ($\epsilon \sim 0.3$) were smaller than the ellipticities of the galaxy isopleths ($\epsilon \sim 0.5$). However, the increasing evidence for substructure in clusters (e.g., West 1995) calls into question any analysis of clusters that approximates them as relaxed systems. Buote & Tsai (1995a; hereafter BT) used the N-body / hydrodynamic simulation of Katz & White (1993) to test the reliability of X-ray constraints of intrinsic cluster shapes considering the effects of substructure. For the X-ray method to be reliable they concluded it is necessary that there is no subclustering on the same scale used to compute the shape of the cluster. Specifically, BT demonstrated that X-ray constraints of the aggregate shape of the Katz & White cluster on scales $r \sim 1.5$ Mpc are very insensitive to subclustering on scales $\lesssim$ few hundred kpc. This is simply a statement that in a hierarchical clustering scenario the bulk of a cluster within $r \sim 1 - 2$ Mpc may be essentially relaxed even though a small subcluster (or two) has recently fallen in to the central regions.

Following the suggestion of BT, in this paper we use data from the ROSAT satellite to probe the aggregate shapes of the bright, low-redshift ($z < 0.1$) Abell clusters A401, A1656 (Coma), A2029, A2199, and A2256 corresponding to clusters we analyzed previously with IPC data (BC; Buote 1992); some of these clusters have now been shown to have significant subclustering in their cores (e.g., Briel et. al. 1991; White, Briel, & Henry 1993; Mohr, Fabricant, & Geller 1993). The image reduction and spatial analysis is described in §2. The hydrostatic modeling of the clusters, the results for intrinsic shapes, and the baryon fractions are discussed in §3. We present our conclusions in §4

## 2. Spatial Analysis of X-ray Data

The cluster images were obtained from the ROSAT Public Data Archive operated by the HEASARC-Legacy database at Goddard Space Flight Center; see Trümper (1983) for a description of ROSAT and Aschenbach (1988) for a discussion of the X-ray telescope. We selected images observed with the Position Sensitive Proportional Counter (PSPC; Pfeffermann et al. 1987) instead of the High Resolution Imager (HRI) because of the PSPC's superior sensitivity, larger field of view, and, most importantly, all of the clusters were observed with the PSPC. Each cluster has at least one observation pointed near the center of the cluster, while a few of the clusters have multiple pointings, some of which are offset from the centers of the clusters. Only certain pointings, however, are useful for our analysis of the intrinsic shapes and mass distributions in the clusters.

Interior to the $40'$ diameter ring of the window support structure of the PSPC the spatial resolution is substantially higher than outside (Hasinger et al. 1994). Outside the ring the PSPC "spokes" supporting the window fan radially outward from the ring and thus add structure to the PSPC images. Since the poor resolution and "spokes" outside the ring degrade and bias measurement of the ellipticity of the surface brightness (i.e. quadrupole; see §2.3) we restrict our analysis to regions interior to the PSPC ring (although see §2.2 regarding A2199). Moreover, when considering multiple pointings, we only used those that completely encircled a cluster about its center. In Table 1 we list the ROSAT sequence numbers and exposure times for the relevant observations of the clusters.

## 2.1. Image Reduction

To prepare the images for analysis we (1) removed time intervals of high background, (2) selected PI bins corresponding to photon energies between 0.5 and 2 keV, (3) corrected for exposure variations and telescopic vignetting, (4) merged multiple pointings for relevant clusters into one image, (5) subtracted the background (only for radial profile), and (6) rebinned the image into pixels corresponding to $\sim 50 h_{80}^{-1}$ kpc ($H_0 = 80 h_{80}$ km s$^{-1}$ Mpc$^{-1}$). All of these reduction procedures were implemented with the standard IRAF-PROS software.

All of the pointed observations were partitioned into many short exposures in order to maximize the observational efficiency of the ROSAT observing program. We examined the background light curves of the images for short-term enhancements indicative of contamination from scattered X-rays, especially from the sun, the bright earth, and the SAA. Only images for A1656 and A2256 required time-filtering. We list the effective exposure times of these filtered pointings in parentheses in Table 1.

To minimize the effects of the X-ray background and the width of the PSPC point spread function (PSF; see §2.2) we selected photons only from energy channels between 0.5 and 2 keV. In addition, we rebinned the images into more manageable $15''$ pixels corresponding to $512 \times 512$ fields. This pixel scale is the same as the exposure maps provided with the standard analysis systems software (SASS); note that the true resolution of the exposure maps actually corresponds to $30''$ pixels.

The images were then flattened using the SASS exposure maps. When dividing the images by these exposure maps, we corrected for both exposure variations and telescopic vignetting. In principle this correction depends on the energy of each individual photon, but for energies above 0.2 keV the energy dependence is small and we neglect it (Snowden et al. 1994).

We merged multiple pointings for A401, A1656, and A2256. Point sources common to the images of each cluster were used to align the fields. A401, which has two images nominally centered on the same position, required a N-S shift of $1.8 \pm 0.5$ pixels, slightly larger than the $\sim 1$ pixel uncertainty (i.e. $15''$) expected of the pointings of ROSAT. The merged images for A1656 included $\sim 500 h_{80}^{-1}$ kpc within the ring of the PSPC. The five pointings for A2256 easily fit 1.5 $h_{80}^{-1}$ Mpc radius within the PSPC ring.

The next step is to remove point sources embedded in the cluster continuum emission. It is imperative to carefully remove such sources to prevent contamination of measurements of the ellipticity of the surface brightness (see BT). Our preferred method to remove sources is by "symmetric substitution" as outlined in BT. This method exploits the property that since our models used for analysis of the intrinsic shape (see §3.1) assume elliptical symmetry we may replace a localized region of the surface brightness with the corresponding regions related by reflection over the major and minor axes of the elliptical isophotes. Unfortunately, this method may only be implemented for an image containing very few sources and having well-defined elliptical symmetry axes. Only A2029 and A2199 meet these requirements for the clusters in our sample. For the other clusters we removed sources by first selecting a source by eye and choosing an annulus around the source to estimate the local background. We then fit a second order polynomial surface to the background and replace the source with the background. This procedure is well suited for estimating quadrupole moments of high $S/N$ cluster images; see Buote & Tsai (1995c) for a thorough discussion.

We estimated the background for each cluster from source-free regions outside the ring of the PSPC fields; these regions were away from the field center at radii $r \sim 40'$ for A401, $30'$ for A2029, $45'$ for A2199, and $55 - 60'$ of the *cluster* center for A2256. The background rates in the 0.5 - 2 keV band of these regions are listed in Table 1. For Coma we adopted the value of White, Briel, & Henry (1993) since they explored regions much farther from the cluster center than we do. We mention that our background estimate for A2256 agrees well with the value $2.24 \times 10^{-4}$ cts s$^{-1}$ arcmin$^{-1}$ of Briel & Henry (1994).

Finally, we rebinned the images of each cluster in order to probe the aggregate cluster structure as outlined in BT. We rebinned the images into pixels corresponding to $\sim 50 h_{80}^{-1}$ kpc for computation of the radial profiles of the clusters; this arbitrary choice provides enough resolution to determine the radial structure suitable for analysis of the aggregate radial structure of the cluster. However, because reliable computation of the ellipticity (see §2.3) requires a sufficiently large number of pixels in a given aperture, we generally selected a finer pixel scale for computing the ellipticities (see Table 2).

Contour plots of the reduced images for each cluster are plotted in Figure 1 using the pixel scales of the radial profiles listed in Table 2. The image for Coma stands out since we are viewing a much smaller region ($r \sim 500 h_{80}^{-1}$ kpc) than for the other clusters ($r \sim 1500 h_{80}^{-1}$ kpc). Binned into these large pixels Coma appears very smooth whereas on smaller scales there is significant structure (see §2.2). This is in keeping with our strategy to measure the aggregate structure of the clusters to reduce possible nonequilibrium effects of small-scale subclustering.

## 2.2. Radial Profile

We constructed the azimuthally averaged radial profile for each cluster using the background-subtracted images. First we computed the centroid of the cluster within a circular aperture containing $\sim 90\%$ of the total flux. This centroid was determined by selecting an origin by eye and then iterating until the centroid changed by $< 0.1$ pixels. For the clusters with known cooling flows (i.e. A2029 and A2199) we excluded the region interior to the radius where the cooling time $\sim$ cluster age; this radius is typically $\sim 100$ kpc (see Sarazin, O'Connell, & McNamara [1992] concerning A2029; see Fabian [1994] for a review of cooling flows in clusters). We exclude the cooling flow region since our models in §3.1 do not apply there.

When computing the centroid of A2256 we must carefully consider the well-known substructure in the core (Briel et al. 1991). Our analysis of the mass distribution in §3.1 assumes the cluster is in hydrostatic equilibrium. We prefer to exclude the region in A2256 were the gas is unrelaxed. To determine where in the cluster the gas is likely to be relaxed we appeal to the power ratios of A2256 (Buote & Tsai 1995b,c). The power ratios classify clusters according to their dynamical state and depend on the scale of the cluster being probed. Buote & Tsai (1995c) find that on a scale of $0.5 h_{80}^{-1}$ Mpc A2256 has power ratios similar to clusters that are manifestly unrelaxed. However, on $1.0 h_{80}^{-1}$ Mpc scales A2256 has power ratios similar to clusters with only a small amount of substructure and is morphologically much closer to evolved clusters like A2029 than on the $0.5 h_{80}^{-1}$ Mpc scale. This effect of more complete relaxation away from the subcluster is supported by the temperature distribution (Briel & Henry 1994) which is complex (with measured azimuthal variations) within $0.5 h_{80}^{-1}$ Mpc and nearly uniform outside $1.0 h_{80}^{-1}$ Mpc. To achieve a balance between eliminating as much of the inner $\sim 0.5 h_{80}^{-1}$ Mpc because of substructure and sampling enough of the radial profile to usefully constrain the mass models in §3.1, we decided to excise the inner $\sim 300 h_{80}^{-1}$ kpc from analysis which effectively encompasses the subcluster in question. We thus compute the centroid in an annular aperture from $(300 - 1500) h_{80}^{-1}$ kpc.

For all of the clusters we computed the radial profiles out to $\sim 1.5 h_{80}^{-1}$ Mpc where the specific outer bin is determined by a $S/N$ criterion (see below). Both Coma and A2199, however, fall well short of $1.5 h_{80}^{-1}$ Mpc within the PSPC ring. We decided to extend the radial profile of A2199 outside the ring in order to compute its baryon fraction to larger distances (see §3.3); the radial profile is not as sensitive as the ellipticity to the larger PSF and spokes outside the ring. Extending the radial profile does not improve our constraints on the intrinsic shape of A2199 since it is the X-ray ellipticity that is the most important determinant which is still confined to within the ring (see Buote & Canizares 1995a). We do not extend the radial profile of Coma since its baryon fraction has been analyzed in detail by others (White et al. 1993).

We rebinned the radial profiles so that all bins had an appropriate minimum signal-to-noise ratio ($S/N$): 7 for A2029 and A2199, and 20 for A2256; A401 and Coma required no additional binning. The background-subtracted radial profiles are displayed in Figure 2. All of the radial profiles appear to be quite smooth for our chosen bin sizes. In particular, Coma's profile displays no large irregularities when binned into $\sim 50h_{80}^{-1}$ kpc pixels, but is known to be very lumpy when probed on galactic scales of $\sim 5$ kpc (e.g., White et al. 1993; Davis & Mushotzky 1993). This property that the cluster is sufficiently relaxed for analysis of its aggregate structure while it is clearly not a smooth, equilibrium configuration on smaller scales exemplifies the arguments of BT and is the driving force for our present investigation.

A convenient parametrization of the X-ray radial profiles of galaxy clusters is the $\beta$ model (Cavaliere & Fusco-Femiano 1976; Jones & Forman 1984; Sarazin 1986),

$$\Sigma_x(R) \propto \left[ 1 + \left( \frac{R}{a_X} \right)^2 \right]^{-3\beta+1/2}, \tag{1}$$

where $a_x$ and $\beta$ are free parameters. The $\beta$ model is especially useful as (1) a benchmark for comparison of $\Sigma_x(R)$ to other clusters and (2) an analytic parametrization of the radial parameters for computing the gas mass. In order to obtain physical constraints on the parameters $a_X$ and $\beta$, we convolved $\Sigma_X$ with the off-axis PSPC PSF (Hasinger et al. 1994) and performed a $\chi^2$ fit to the radial profile; note for the evaluation of the PSF we set the energy to 1 keV and for A1656 and A2256, which have pointings on different regions of the clusters, we fixed the off-axis angle to $10'$. Since our radial bins are larger, in some instances substantially larger, than the width of the PSF, including the PSF in the fits has a negligible ($\lesssim 1\%$) effect on the fitted parameters. Nevertheless, we include the PSF in the fits for completeness.

In Table 3 we list the best-fit parameters and 95% confidence limits on two interesting parameters; the best-fit models are also plotted in Figure 2. The $\beta$ model is an excellent visual fit to the radial profiles for all of the clusters, although the $\chi^2$ values are rather large. For the purposes of analyzing the aggregate structure of the clusters on $\sim 1$ Mpc scales, though, the fits give very acceptable descriptions of the surface brightness distribution. Except for A2256, the fitted parameters are generally insensitive to the aperture size used to evaluate the centroid; i.e. $\Delta a_x \lesssim 5\%, \Delta \beta \lesssim 1\%$. A2256, which is known to have a large centroid shift (Fabricant, Rybicki, & Gorenstein 1984; Mohr et al. 1993), has substantially different parameters depending on the chosen aperture used for computing the centroid. The values listed in Table 3 for A2256 are for the centroid evaluated in an annular aperture $(300 - 1100)\ h_{80}^{-1}$ kpc representing $\sim 90\%$ of the flux in the annulus $(300 - 1500)\ h_{80}^{-1}$ kpc. If we instead use a circular aperture of radius $1500h_{80}^{-1}$ kpc then we obtain best-fit parameters $a_x = 5'.21$ and $\beta = 0.815$. These values are in excellent agreement with those obtained by Briel & Henry (1994) who removed the effects of the subcluster by excising a pie slice from the image with origin located at the primary cluster center and edges extending throughout the whole cluster. Our fitted $\beta$-model parameters when including the subcluster agree with those of Briel & Henry when they exclude the subcluster in a pie slice; i.e. excluding the subcluster itself has a minimal effect on the radial profile but when the whole interior is excluded the $\beta$-model parameters are noticeably affected. This may simply reflect the steepening of the radial profile outside the core. As stated above, we prefer to exclude from analysis the region $r < 300 h_{80}^{-1}$ kpc because the substructure there may signal substantial departures from hydrostatic equilibrium. Hence, except for A2256, the clusters have $a_x$ and $\beta$ that agree well with those found in the literature (e.g., Jones & Forman 1984; Briel, Henry, & Böhringer 1991).

### 2.3. Ellipticity

We quantify the elongation of the clusters using the iterative moment technique introduced by Carter & Metcalfe (1980). This method is particularly suited to measure the aggregate ellipticity of the clusters in a large aperture (see BT and Buote & Canizares [1994] for detailed discussions). In essence, this technique entails computing the analogue of the two-dimensional moments of inertia arrived at by iterating an initially circular region; the square root of the ratio of the principal moments is the axial ratio and the orientation of the principal moments yields the position angle. The parameters obtained from this method, $\epsilon_M$ and $\theta_M$, are good estimates of the ellipticity ($\epsilon$) and position angle ($\theta$) of an intrinsic elliptical distribution of constant shape and orientation. For a more complex intrinsic shape distribution, $\epsilon_M$ and $\theta_M$ are average values weighted heavily by the outer parts of the region.

We applied a simple Monte Carlo procedure to characterize the uncertainties on $\epsilon_M$ and $\theta_M$ due to undetected point sources and Poisson noise. For each cluster we simulated $\beta$ models having the best-fit $a_x$ and $\beta$ obtained in §2.2 modified to have a constant ellipticity and orientation; i.e. we replaced $R = \sqrt{x^2 + y^2}$ with the elliptical radius $\sqrt{x^2 + y^2/q^2}$, where $q$ is the constant axial ratio. These models were scaled to have the same number of counts (or count rate) as the background-subtracted PSPC images of the clusters. We then added the uniform background for each cluster (see Table 1). To each image we added point sources having spatial properties consistent with the PSPC PSF and numbers consistent with the $\log N(> S)$ - $\log S$ distribution given by Hasinger (1992); see Buote & Tsai (1995b) for a detailed discussion. Finally, we added Poisson noise to the composite images. Since $\epsilon_M$ is unaffected by a uniform background (see Carter & Metcalfe) we did not then subtract the background. We performed 1000 simulations each for a suite of input ellipticities ($\epsilon_x = 1 - q$) for each cluster.

To determine the 95% confidence intervals on the measured $\epsilon_M$ we proceeded as follows; for now we focus our attention on a particular aperture size. We arrange the results of the 1000 simulations for a given input $\epsilon_x$ into ascending order of measured ellipticity $\epsilon_M$; i.e. $\epsilon_M^1 < \epsilon_M^2 < \cdots < \epsilon_M^{1000}$. The 95% upper limit for this model is defined to be the value of $\epsilon_M$ corresponding to the $0.95 \times 1000 = 950$th value of $\epsilon_M$ in the ordered array; i.e. $\epsilon_M^{950}$. The 95% confidence lower limit of the real data is given by the model with input $\epsilon_x$ whose $\epsilon_M^{950}$ just equals the measured value of $\epsilon_M$ from the real image (see below). Similarly, the 95% confidence upper limit of the real data is given by the model with input $\epsilon_x$ whose $\epsilon_M^{50}$ just equals the measured value of $\epsilon_M$ from the real image.

We estimated the confidence limits for the position angles $\theta_M$ from the model having $\epsilon_x = \epsilon_M$ of the data for the specific aperture size. We define the 95% confidence limits to be $(\theta_M^{25}, \theta_M^{975})$. This procedure is not the most rigorous means to determine the uncertainties on $\theta_M$ because the value of $\epsilon_M$ in each simulation is not equal to $\epsilon_x$. Since, however, we do not use $\theta_M$ in our modeling (see §3.1) the estimates are sufficient for our purposes.

In Table 4 we list the values of $\epsilon_M$ and $\theta_M$ and their associated 95% confidence limits for different apertures (in 5 pixel increments) for each of the clusters. Similar to our analysis of the radial profiles, we excluded the cooling-flow regions of A2029 and A2199 from the $\epsilon_M$ computation. Similarly, we excluded the region $a \lesssim 300 h_{80}^{-1}$ kpc for A2256 to lessen the influence of the core substructure. Generally the values of $\epsilon_M$ are larger than the ellipticities obtained by BC (Buote 1992) from *Einstein* IPC images of these clusters; note BC used a slightly modified version of the iterative moment technique. Other published ellipticities for these clusters using IPC data are consistent with the results of BC (Buote 1992). For example, using different techniques to measure the ellipticity Fabricant et al. (1984), Buote (1992), Mohr et al. (1993), and Davis & Mushotzky (1993) measured an ellipticity of $\sim 0.2$ for A2256 at $10'$ – all significantly less than the value of $\sim 0.3$ computed in this paper. For Coma BC and Mohr et al. obtained ellipticities at $15'$ in good agreement with the value of 0.16 obtained in this paper although Davis & Mushotzky obtained $\sim 0.27$. The good agreement between IPC and PSPC results for Coma is expected since the effect of the PSF should be least important for Coma of all the clusters in our sample. Hence, the discrepancy between the ellipticities of the clusters computed with IPC data and the results of the present paper reflects the superior spatial resolution of the PSPC.

All of the clusters except A1656 have $\epsilon_M$ that are consistent within their uncertainties with being monotonically decreasing with increasing aperture size; in fact, the ellipticities of Coma appear to be increasing with radius but over a much smaller range. The position angles vary $\lesssim 10°$ over the regions probed except for Coma and A2029 which have $\sim 15°$ variations. The $\theta_M$ are generally in good agreement with the previous IPC studies and with the orientations of the galaxy isopleths (Carter & Metcalf).

Following BT we designated two apertures within the $\sim 1.5 h_{80}^{-1}$ Mpc regions for use in constraining the mass models in §3.1; for Coma we only used the largest aperture available. Using two large, well-separated apertures is in keeping with our scheme to analyze the aggregate structure of clusters while still allowing us to obtain information on any ellipticity gradients. We denote these special apertures with an asterisk in Table 4.

## 3. Intrinsic Shapes and Mass Profiles from the X-rays

### 3.1. Method

The technique we use to constrain the intrinsic shape and mass profile of the total gravitating mass for a cluster of galaxies is discussed in detail by BT. Buote & Canizares (1994,1995a) describe how we use the the observed light distributions to infer the shape and profile of the dark matter from X-ray images. We refer the reader to these papers for exposition of the modeling procedures we employ in this paper. Here we summarize only a few small additions to the method introduced for this particular study.

Because of the increasing evidence from gravitational-lens observations of giant arcs (e.g., Miralda-Escudé & Babul 1995) and N-body simulations (e.g., Navarro, Frenk, & White 1995) that galaxy clusters have small or perhaps no core radii, in this paper we will explore density profiles that are simple generalizations of the Hernquist (1990) model,

$$\rho_H \propto (am)^{-1} \left(R_c + am\right)^{-n},\tag{2}$$

where $R_c$ is a core parameter, $a$ is the semi-major axis of the bounding spheroid of the SMD (Spheroidal Mass Distribution – see BT), and $m$ is the dimensionless spheroidal radius. For $am \ll R_c$, $\rho_H \sim r^{-1}$, and thus these densities have no core. At large radii, $am \gg R_c$, $\rho_H \sim r^{-(n+1)}$ and falls as $r^{-4}$ for the Hernquist (1990) density ($n = 3$) we used in previous studies. We consider models having $n$ ranging from $1 - 3$ which covers plausible ranges of density profiles of clusters. These models complement the "cored" power-law SMD models ($\rho \propto [R_c^2 + (am)^2]^{-\eta}$) that we also use to model the masses of clusters.

To describe the mass distribution of the X-ray-emitting gas we use simple models having isodensity surfaces that change shape with radius. We are lead to this choice since the X-ray ellipticity profiles of the clusters measured from the PSPC data (Table 4) display significant gradients. An accurate model for the gas distribution is desirable not only for obtaining a precise measure of the total gas mass, but also for constraining the shape and profile of the dark matter. That is, in galaxy clusters the dominant mass component is the dark matter which typically contains $80\% - 90\%$ of the mass. The gas mass is $10\% - 20\%$ of the total while the mass of the stellar material, which is not nearly as well constrained as the gas, is generally much smaller than the gas mass (e.g., the case of A2256 – Briel & Henry 1994). Since the total mass of the galaxies is not well constrained, and in any case probably insignificant with respect to the gas, we neglect it (i.e. $M_{tot} = M_{gas} + M_{DM}$); note that for elliptical galaxies the opposite is true – the gas may be neglected in favor of the stellar mass (e.g., Buote & Canizares 1994,1995a).

The models we use to describe the gas mass are the oblate (and prolate, see below) spheroids of varying ellipticity introduced by Ryden (1990). Ryden computes the monopole and quadrupole terms of the gravitational potential generated by a mass distribution of concentric, oblate, spheroidal shells whose eccentricity ($e^2 = 1 - q^2$) varies as a function of the semi-major axis of the shell. Like Ryden, we consider spheroids that have eccentricity,

$$e(a) = \frac{e_0}{1 + (a/a_0)^\alpha},$$

where we have inserted the extra parameter $e_0$ to allow for an arbitrary central eccentricity. This simple function yields surprisingly good descriptions of the ellipticity profiles of the X-ray isophotes (see Table §4). Since we also desire to explore prolate models we extend Ryden's formalism to the prolate case in the Appendix.

For each cluster we construct Ryden models for the gas mass. Since $\beta$ models describe the radial profiles of the clusters in our sample so well (§2.2) we model the three-dimensional density profiles as $\rho_{gas} = (a_x^2 + a^2)^{-3\beta/2}$. Here $a_x$ and $\beta$ are the best-fit parameters obtained from fitting the $\beta$ model convolved with the PSPC PSF to all of the radial bins for each cluster. To completely specify the gas model we assign $e(a)$ (eqn. [3]) to $\rho_{gas}$ appropriate to the ellipticity of the X-rays. That is, we generate a model X-ray surface brightness by projecting $\rho_{gas}^2$ onto the sky plane and convolving it with the PSPC PSF. By comparing the ellipticities of this model with those of the real image (see Table 4) we determine the parameters of $e(a)$.

## 3.2.    Shapes of the Composite Mass Distributions

Our procedure to constrain the hydrostatic models using the radial profiles (§2.2) and ellipticities (asterisked values in Table 4) is described in detail in BT §4.2. For all of the spheroidal models we set the symmetry axis to lie in the plane of the sky; i.e. we are not attempting to uncover projection effects in this analysis, although we believe them to be small for these clusters (see §3.4). We set the semi-major axis of the SMD models to $1.5h_{80}^{-1}$ Mpc for all the clusters.

### 3.2.1.    Isothermal Models

Recent ASCA observations show that that gas is nearly isothermal within the central $\sim 1$ Mpc of rich clusters (Mushotzky 1995). Since the shape of a cluster inferred from X-ray analysis is mostly insensitive to small temperature gradients (BT; also Strimple & Binney 1979 and Buote & Canizares 1994), we focus our attention on isothermal models. The effects of temperature gradients consistent with PSPC data are discussed in §3.2.2.

We list in Table 5 the 95% confidence limits on the ellipticities of the composite mass ($\epsilon_{mass}$) for the isothermal models that best characterize the X-ray data. Similar to our experience from fitting the $\beta$ models (§2.2), the isothermal models give excellent visual fits to the data but with rather large values of $\chi^2$. The two cooling-flow clusters, A2029 and A2199, require steep mass density profiles ($\rho_{mass} \sim r^{-4}$) to reproduce the observed ellipticity gradients of the X-rays; even with this steep density the models for A2029 only marginally reproduce the X-ray ellipticity gradient within the estimated 95% confidence uncertainties. The ellipticities in Table 5 for A2029 and A2199 are for the Hernquist (1990) models (i.e. $n = 3$ in eq.[2]), although the cored power-law models with $\eta = 2$ give essentially identical values.

The clusters A401 and A2256, which have no observed cooling flows (Fabian 1994), also have steep X-ray ellipticity gradients that imply steep mass density profiles ($\rho_{mass} \sim r^{-4}$). For these clusters, though, we list the results in Table 5 for the cored power-law models with $\eta = 2$ since their radial profiles have large core radii (§2.2); as was the case for A2029 and A2199, the ellipticities for A401 and A2256 are not very sensitive to the density model and we obtain similar results for the Hernquist models. Coma, for which we only use a single aperture to constrain the ellipticity (and thus have no information on the large-scale ellipticity gradient), is fit equally well by all of the models considered. We show in Table 5 the results for both the $\eta = 1, 2$ cored power-law models ($\rho_{mass} \sim r^{-2}$ and $\rho_{mass} \sim r^{-4}$).

The ellipticities required by the isothermal models are quite large for all of the clusters, much larger than those we obtained from analysis of *Einstein* IPC data in BC. However, the ellipticities for A2256 are in good agreement with IPC analyses of Fabricant, et al. (1984) and Buote (1992). We attribute the discrepancy to BC not accounting for the steep ellipticity gradients of the X-ray isophotes obtained for these clusters (see Table 4). In BC we selected only one aperture size to constrain the X-ray ellipticity and the size of the aperture was only a few times the width of the IPC PSF for all the clusters but Coma (and Perseus). Since in BC we did not incorporate the IPC PSF into the hydrostatic modeling procedure we did not accurately estimate the uncertainty due to possible ellipticity gradients being smeared out by the IPC PSF. In the case of Coma the discrepancy is due primarily to the larger aperture we use in this paper to compute the X-ray ellipticity; when we use the same aperture size as in BC we obtain $\epsilon_{mass}$ in good agreement. The agreement of the shape of A2256 with previous analyses appears to have been due to a fortuitous conspiracy of the different region analyzed in this paper and the larger X-ray ellipticities due to the better spatial resolution of the PSPC.

### 3.2.2. Models with Temperature Gradients

We now consider temperature gradients allowed by the PSPC data. Unfortunately, since the PSPC energy band corresponds to much lower energies than the temperatures of the rich clusters in our sample we are unable to obtain precise constraints on the temperature gradients. In general we find that the uncertainty in the gradients allowed by the PSPC data dwarf those obtained by PV-phase ASCA observations of rich clusters (Mushotzky 1995). Instead of giving serious consideration to models that do not appear to be realistic, we will simply illustrate this uncertainty with the case of A2029. We defer serious consideration of the effects of real temperature gradients in these clusters on the shapes to analysis of ASCA data (Buote & Canizares 1995b).

In order to obtain the narrowest constraints on the temperature gradient we placed plausible restrictions on the spectrum and the temperature profile. Our intention is to demonstrate that even when such restrictions are applied there is still an implausibly large range of temperature gradients allowed. We first divide the surface brightness of A2029 into two annular regions having equal $S/N$: (140-336,336-1400) $h_{80}^{-1}$ kpc; we exclude the region $r < 140 h_{80}^{-1}$ kpc because of the cooling flow. The temperature of each region was determined by fitting (with XSPEC) a Raymond-Smith (1977) model with Galactic absorption and 50% solar metallicities. We obtain $T_{in} = 7.7(6.7 - 9.3)$ keV and $T_{out} = 6.1(4.8 - 8.1)$ keV at 68% confidence.

We take the three-dimensional temperature profile to follow a broken power law,

$$T(r) = T_0 \qquad\qquad r \leq r_0 \qquad (4)$$

$$T(r) = T_0 \left(\frac{r}{r_0}\right)^p \qquad r > r_0 \qquad (5)$$

where $r_0 \equiv 140 h_{80}^{-1}$ kpc, and $T_0$ and $p$ are free parameters. For the purposes of the determination of the temperature gradients implied by the spectral constraints, we suppress the information supplied by the elongation of the X-ray isophotes which only has a small effect on the parameters $T_0$ and $p$. However, for the hydrostatic models (see below) we force the isotemperature surfaces given by equation [5] to be stratified on the isopotential surfaces.

We construct the emission-weighted temperature ($\langle T \rangle$) projected along the line of sight ($dl$) into an area $dA$,

$$\langle T \rangle = \frac{\int dA \int_{-\infty}^{\infty} j_{gas} T \, dl}{\int \Sigma_x \, dA}, \qquad (6)$$

where $j_{gas}$ is the gas volume emissivity and $\Sigma_x = \int_{-\infty}^{\infty} j_{gas} \, dl$ is the surface brightness. To determine $T_0$ and $p$ we evaluate equation[6] over $A$ corresponding to the two annular regions (140-336,336-1400) $h_{80}^{-1}$ kpc and then perform a $\chi^2$ fit to the 68% confidence limits on the temperatures; note we defined the temperatures in the bins to be the middle of the 68% confidence temperature results, and the $1\sigma$ weights are half the widths of the confidence intervals.

The best-fit parameters are $T_0 = 8.8$ keV and $p = -0.27$; the 68% lower limit is $T_0 = 11.1$ keV and $p = -0.68$ and the upper limit is $T_0 = 6.7$ keV and $p = 0.07$. The slopes of the best-fit ($p = -0.27$) and 68% lower limit ($p = -0.65$) temperatures imply substantially larger gradients than observed in real clusters (Mushotzky 1995), although the 68% upper limit slope is modest ($p = +0.07$). We list in Table 6 the derived mass ellipticities for the temperature gradients determined above using the same mass density models for A2029 given in Table 5. As expected, the ellipticities for the small gradient given by the 68% upper limit are shifted down only by $\sim 0.03$ with respect to the isothermal results; even $\epsilon_{mass}$ for the best-fit temperature parameters are shifted upwards by only $\sim 0.05$. However, the very large gradient given by the 68% lower limit temperature gradient implies implausibly flat mass distributions for a non-rotating, self-gravitating spheroid (Merritt & Stiavelli 1990; Merritt & Hernquist 1991).

In the previous discussion we have only considered the effects of radial temperature gradients on the derived shape of the total gravitating cluster mass. Azimuthal temperature variations can distort the shapes of the X-ray isophotes since the X-ray emissivity also depends on the gas temperature via the plasma emissivity (convolved with the spectral response of the detector). Since the PSPC energy pass band corresponds to smaller energies than the typical temperatures of rich clusters the plasma emissivity convolved with the spectral response of the PSPC is effectively constant for the clusters in our sample (see NRA 91-OSSA-3, Appendix F, ROSAT mission description); i.e. the isophote shapes measured with the PSPC are largely insensitive to temperature variations. Of more serious concern is that large azimuthal temperature gradients likely imply strong departures from hydrostatic equilibrium in which case a non-equilibrium analysis is required; e.g., the temperature distribution of the core of A2256 has been modeled using hydrodynamic simulations (Burns 1995). However, such clusters often exhibit substructure which allows one to judge (without knowing the temperature) where in the cluster hydrostatic analysis is appropriate (as in A2256 – see §2.2).

### 3.3. The Total Gravitating Mass and Baryon Fraction

In order to determine the shapes of the dark matter we must first know the ratio of the gas mass to total gravitating mass (see §3.1) for the clusters. To compute the gas mass we simply integrated the models for $\rho_{gas}$ in §3.1. We normalized $\rho_{gas}$ for each cluster by first constructing the X-ray emissivities, $j_{gas}$, (see eq.[9] of Buote & Canizares 1994) from $\rho_{gas}$ and then projected $j_{gas}$ onto the sky plane to obtain $\Sigma_x$. We then normalized $\Sigma_x$ to the total flux between 0.5 - 2.0 keV determined by fitting the spectrum (with PROS) to a Raymond-Smith (1977, updated to 1992 version) model with Galactic absorption (David et al. 1993 who use the results of Stark et al. 1992) and temperatures from Edge et al. (1990); see Table 2. By normalizing to the flux we have completely specified $\rho_{gas}$ and hence the gas mass. Our estimates for the gas mass take into account the following uncertainties: (1) 95% confidence statistical errors on the $\beta$-model parameters (see Table §3) used to define the radial profile of $\rho_{gas}$, (2) oblate and prolate geometry, and (3) variation in flux and plasma emissivity due to the 90% uncertainties in the gas temperatures from Edge et al.. There is an additional source of uncertainty due to the restriction of the adopted models for $\rho_{gas}$ to having radial profiles of the $\beta$ models. Since, however, the $\beta$ models describe the clusters so well over the regions considered the uncertainty in the gas mass should be quite small, certainly $\lesssim 10\%$. We list in Table 7 the gas masses computed in spheres of radii $(0.5, 1.0, 1.5)h_{80}^{-1}$ Mpc considering (1) - (3) for all of the clusters; note for Coma we only compute the gas mass out to $0.5h_{80}^{-1}$ Mpc.

The gas masses are extremely well constrained for the clusters, generally to better than a few percent. Our values for $M_{gas}$ appear to be slightly larger than estimates found in the literature. White & Fabian (1995) used *Einstein* IPC data to compute gas masses and total masses of clusters assuming spherical symmetry for a sample of Abell clusters, including A401 and A2029. They obtain $M_{gas} = 1.32 \pm 0.07 \times 10^{14} M_{\odot}$ for A401 ($r \leq 1.265$ Mpc) and $M_{gas} = 1.26 \pm 0.11 \times 10^{14} M_{\odot}$ for A2029 ($r \leq 1.291$ Mpc), all quantities evaluated for $H_0 = 50$ km s$^{-1}$ Mpc$^{-1}$; note that $M_{gas} \propto h^{-5/2}$. Using the same apertures and Hubble constant as White & Fabian we obtain $M_{gas} = (1.99 - 2.02) \times 10^{14} M_\odot$ for A401 and $M_{gas} = (1.90 - 1.97) \times 10^{14} M_\odot$ for A2029. Although our values are only $\sim 50\%$ larger, the discrepancy appears to be significant within the estimated uncertainties. White & Fabian, however, acknowledge that their estimates of the gas masses are likely to be somewhat conservative because of the procedure they employ to obtain core radii for the gas profiles. If we neglect the ellipticity of the surface brightness the masses are hardly affected; e.g., $M_{gas} = (1.85 - 1.92) \times 10^{14} M_\odot$ for A2029. This confirms earlier statements that ellipticity has negligible impact on estimates of gas masses in clusters (e.g., White et al. 1994). By extrapolating our model for $\rho_{gas}$ to a 3 Mpc ($H_0 = 50$ km s$^{-1}$ Mpc$^{-1}$) radius for A2029 we obtain $M_{gas} = (5.6 - 5.9) \times 10^{14} M_\odot$, in good agreement with Jones & Forman's (1984) value of $M_{gas} = 5.3 \times 10^{14} M_\odot$ obtained from IPC data considering their expected (unstated) uncertainties.

The gas mass for A2256 has been computed by Briel & Henry (1994) using the ROSAT PSPC data. They obtain $M_{gas} = 6.33 \pm 1.17 \times 10^{14} M_\odot$ within a 1.4 Mpc radius ($H_0 = 100$ km s$^{-1}$ Mpc$^{-1}$). At this distance we obtain (neglecting the ellipticity of the gas) $M_{gas} = (7.46 - 7.84) \times 10^{14} M_\odot$, only $\sim 15\%$ larger than the result of Briel & Henry. This small discrepancy can likely be attributed to that fact that Briel & Henry exclude a large section of the data from analysis different from us (see §2.2) in order to avoid the subclump in the interior.

Using the isothermal spheroidal models of Table 5 we present in Table 7 the total gravitating masses of the clusters computed in spheres of radii $(0.5, 1.0, 1.5) h_{80}^{-1}$ Mpc. The uncertainties in the masses reflect 95% confidence statistical errors on $R_c$ and $n$ (or $\eta$) from the mass models and the 90% errors on the temperatures from Edge et al. which is the dominant source of uncertainty. It is worth mentioning that because our mass models have steep density profiles (i.e. $\rho \sim r^{-4}$) the results of Table 7 differ from the total masses obtained assuming the cluster is a $\beta$ model; i.e. a $\beta$ model corresponds to a logarithmic potential (e.g., Appendix C of Trinchieri, Fabbiano, & Canizares 1986) and thus $\rho \sim r^{-2}$ at large radii. The difference between models is highlighted by the case of A2199 for which Mushotzky (1995) obtains $M_{tot} = (1.5 - 1.7) h_{80}^{-1} \times 10^{14} M_\odot$ within $0.5 h_{80}^{-1}$ Mpc using the $\beta$ model – we obtain the same result using the derived $\beta$-model parameters in Table 3. However, using the $\rho \sim r^{-4}$ models we obtain (in Table 7) a mass that is over a factor of six(!) larger than the $\beta$ model at $0.5 h_{80}^{-1}$ Mpc, although the descrepancy between the models decreases with increasing radius. However, the discrepancy between models is not so pronounced for all the clusters, especially at large radii. When scaled to 1.4 Mpc ($H_0 = 100$ km s$^{-1}$ Mpc$^{-1}$), we obtain $M_{tot} = (12 - 21) \times 10^{14} M_\odot$ for A2256 using the $\rho \sim r^{-4}$ models, only slighly larger than the values we obtain assuming a $\beta$ model $M_{tot} = (7.0 - 10.8) \times 10^{14} M_\odot$. The $\beta$-model values also happen to be in excellent agreement with the result of Briel & Henry (1994), $\sim 9.5 \times h^{-1} 10^{14} M_\odot$, who also used the isothermal $\beta$-model to compute the mass.

As expected, we find that the masses derived assuming spherical symmetry agree very well with those incorporating the ellipticity of the X-ray isophotes; the viability of X-ray mass estimates assuming spherical symmetry has already been addressed in the literature (Tsai, Katz, & Bertschinger 1994; Navarro et al. 1995). All of the clusters in our sample that are measured to large radii have $M_{gas}/M_{tot}$ increasing with radius. Scaling to $H_0 = 50$ km s$^{-1}$ Mpc$^{-1}$ the mass ratios are $(8-22)\%$ consistent with the results of White & Fabian (1995) for their sample of 19 Abell clusters; $M_{gas}/M_{tot} \propto h^{-3/2}$.

### 3.4. Shapes of the Dark Matter Distributions

Using the ratios $M_{gas}/M_{tot}$ in Table 7 we determine the ellipticity of the dark matter distributions, $\epsilon_{DM}$, for the clusters. In order to consider the maximum possible effects of the self-gravitation of the gas we also consider the upper limits of $M_{gas}/M_{tot}$ scaled to $H_0 = 50$ km s$^{-1}$ Mpc$^{-1}$; i.e. this scaling amounts to doubling the values of $M_{gas}/M_{tot}$ in Table 7. The results for $\epsilon_{DM}$ corresponding to the mass models in §3.2.1 (see Table 5) are listed in Table 8. The ellipticities of the dark matter are hardly affected upon consideration of the self-gravitation of the gas; i.e. for all the clusters $\epsilon_{DM} \approx \epsilon_{tot} + 0.02$.

Also listed in Table 8 are the $1\sigma$ ellipticities of the galaxy isopleths, $\epsilon_{gal}$, computed by Carter & Metcalfe (1980) using the same iterative moment technique of §2.3. All of clusters have $\epsilon_{DM}$ consistent with $\epsilon_{gal}$ within the $(1-2)\sigma$ uncertainties of $\epsilon_{gal}$, although $\epsilon_{DM}$ for A2256 is only marginally consistent with the $2\sigma$ lower limit for $\epsilon_{gal}$. The case for A2256 is actually uncertain since Fabricant, Kent, & Kurtz (1989) obtained a much smaller lower limit for $\epsilon_{gal} = 0.23$ (90% confidence) using a different sample with more redshifts and a different technique to compute the ellipticity from Carter & Metcalf. Hence, the shapes of the dark matter distributions and galaxy isopleths are consistent for each of the clusters in our sample contrary to our previous conclusions in BC. (We mention that $\epsilon_{gal}$ for A1656, A2029, and A2199 obtained by Plionis, Barrow, & Frenk [1991] using a different technique also agrees with $\epsilon_{DM}$. Since Plionis et al. find that the intrinsic $\epsilon_{gal}$ distribution for clusters peaks at $\sim 0.5$, the clusters in our sample are likely viewed nearly edge-on and thus our conclusions for the dark-matter shapes may be typical.)

Finally, the ellipticities of the dark matter distributions for the clusters are consistent with the predictions from dissipationless formation of halos in a universe filled with Cold Dark Matter (Frenk et al. 1988; Efstathiou et al. 1988). The dark matter halos produced by these N-body simulations are triaxial and generally have $\epsilon_{DM} \leq 0.5$ but several have $\epsilon_{DM}$ approaching 0.67. Our mass models fit the observed X-ray radial profiles and ellipticities equally well for both oblate and prolate models which is consistent with the halos being typically triaxial. These results are also consistent with the cluster formed in the simulation of Katz & White (1993) which has $\epsilon_{DM} = 0.54$ within $1.5h_{50}^{-1}$ Mpc (see BT).

### 4. Conclusions

We have analyzed ROSAT PSPC X-ray images of five bright, low-redshift ($z < 0.1$) Abell clusters for the purpose of constraining their intrinsic shapes and mass profiles. The intrinsic shapes of the clusters in our sample (A401,A1656,A2029,A2199,A2256) were analyzed previously by us (Buote & Canizares 1992; Buote 1992) using *Einstein* IPC X-ray images. In this paper we specifically follow Buote & Tsai's (1995a) procedure to constrain the aggregate shapes of clusters on large scales ($\sim 1.5$ Mpc) to reduce the effects of possible substructure on small scales ($\lesssim$ few hundred kpc).

We computed the azimuthally averaged radial profiles and ellipticities (i.e. quadrupole moments) of the X-ray surface brightness distributions within $\sim 1.5 h_{80}^{-1}$ Mpc for the all clusters except Coma where we were restricted to $\sim 0.5 h_{80}^{-1}$ Mpc. Fitting $\beta$ models to the radial profiles yields very precise constraints on the core radii and $\beta$ parameters in good agreement with previous results from *Einstein* (e.g., Jones & Forman 1984). The ellipticities of the X-ray images are also tightly constrained in their 95% confidence intervals and are systematically larger ($\epsilon \approx 0.15 - 0.25$) than those ($\epsilon \approx 0.10 - 0.20$) obtained by Buote & Canizares (1992; Buote 1992) using *Einstein* data. This discrepancy is a direct result of the superior spatial resolution of the PSPC. All of the clusters (except Coma) display substantial ellipticity gradients within $r \sim 1.5 h_{80}^{-1}$ Mpc not seen in the *Einstein* data.

Using the X-ray radial profiles and ellipticities and assuming hydrostatic equilibrium we constrained the ellipticity of the total gravitating matter following Buote & Tsai (1995a). Isothermal mass models for the clusters yield ellipticities, $\epsilon_{mass} \approx 0.40 - 0.55$ (95% confidence), which are systematically larger (i.e. $\Delta\epsilon_{mass} \sim 0.15$) than obtained by Buote & Canizares (1992; Buote 1992) using *Einstein* data; we understand this discrepancy to be a combination of effects due to different cluster regions being probed, modeling differences between the two investigations, and most importantly the X-ray ellipticity gradients obtained from the PSPC data. We mention that the measured X-ray ellipticity gradients require steep mass profiles ($\rho \sim r^{-4}$) for our SMD models; i.e. models where the mass is stratified on concentric spheroids of constant ellipticity. Of course, this does not exclude the possibility of a flatter profile where the mass changes shape with radius. Models with small temperature gradients typical for rich clusters (Mushotzky 1995) affect the isothermal $\epsilon_{mass}$ estimates by $< 10\%$.

We computed the masses of the X-ray–emitting gas and the total gravitating matter in spheres of radii $(0.5, 1.0, 1.5) h_{80}^{-1}$ Mpc using the 95% statistical uncertainties of our gas and total mass models and the 90% uncertainties on the gas temperatures from Edge et al. (1990). The effects of the X-ray ellipticity on the gas masses is generally less than a few percent in agreement with the result of White et al. (1994). The ratio of gas mass to total gravitating mass of the clusters increases with radius and has values $M_{gas}/M_{tot} = (4\% - 11\%) h_{80}^{-3/2}$ within a $1.5 h_{80}^{-1}$ Mpc radius, in excellent agreement with the results of White & Fabian (1995) and thus consistent with the "Baryon Catastrophe" proposed by White et al. (1993).

The shapes derived for the dark matter distributions using the isothermal models are essentially identical to those obtained for the total gravitating matter: $\epsilon_{DM} \approx \epsilon_{tot} + 0.02$. The observed ellipticities of the galaxy isopleths (Carter & Metcalf 1980) are consistent with the ellipticities of the dark matter. Moreover, the ellipticities of dark halos predicted by N-body simulations of a universe dominated by cold dark matter (Frenk et al. 1988; Efstathiou et al. 1988; Katz & White 1993) agree with the results for the clusters in our sample.

Our modeling procedure has generated tight constraints on the intrinsic ellipticities and density profiles of the dark matter from the X-ray data. To obtain the most general constraints a non-parametric estimation of the shape of the dark matter is warranted and should be feasible given the high $S/N$ and spatial resolution of the observations along with the highly relaxed appearance of A401, A2029, and A2199 (see Buote & Tsai 1995b for a discussion of judging degree of relaxation). For these clusters a non-parametric treatment simply involves solving the hydrostatic equation for the potential in terms of the gas density and temperature which may be obtained from general spheroidal deprojection following Palmer (1994).

It is a pleasure to thank E. Bertschinger and J. Tonry for insightful discussions and the anonymous referee for useful suggestions and for being prompt. We extend our gratitude to R. Mushotzky for communicating unpublished ASCA results on the temperature profiles in clusters, M. Corcoran for assistance with the ROSAT archive, D. Harris for his advice regarding merging PSPC images, and to U. Hwang for providing software to compute plasma emissivities. We acknowledge use of the following astrophysical databases: ADS, HEASARC-Legacy, NED, and SIMBAD. This research was supported by grants NAG5-1656, NAS8-38249 and NASGW-2681 (through subcontract SVSV2-62002 from the Smithsonian Astrophysical Observatory).

## A.  Prolate Ryden Potentials

Ryden (1990) computes the gravitational potentials (up to quadrupole order) of oblate spheroids having ellipticity that varies with radius. Generalization of Ryden's result to prolate spheroids is a straightforward (albeit tedious) application of the multipole expansion of the three-dimensional gravitational potential (e.g., Binney & Tremaine 1987). Ryden considers the three-dimensional density distribution to be completely specified by the density, $\rho(a)$, along the semi-major axis, $a$, and the eccentricity, $e(a)$, of the isodensity surfaces. The isodensity surface for the prolate case is,

$$\frac{r^2 \cos^2 \theta}{a^2} + \frac{r^2 \sin^2 \theta}{a^2(1 - e^2)} = 1, \tag{A1}$$

where $r$ is the distance from the center, $\theta$ is the conventional polar angle, and we have suppressed the dependence on $a$ in $e$ for notational convenience. Expressing the gradient in eccentricity as,

$$\Sigma = \frac{a}{1 - e^2} \frac{d(1 - e^2)}{da}, \tag{A2}$$

the mass of a thin shell of matter of width $\delta a$ and uniform density $\rho$ is,

$$\delta M = 4\pi(1 - e^2)\rho a^2 \delta a \left[1 + \Sigma/3\right]. \tag{A3}$$

The oblate potential is completely specified by the coefficients for the interior and exterior potential given by Ryden's equations (7) and (12). For the prolate case we obtain the following values,

$$B_{00} = 8\pi\sqrt{\pi}G\rho a(1-e^2)(1+\Sigma/3) \tag{A4}$$

$$B_{20} = \frac{8\pi\sqrt{\pi}}{3\sqrt{5}}G\rho a e^2(1-e^2)\left[1+\frac{3-2e^2}{5e^2}\Sigma\right] \tag{A5}$$

$$C_{00} = 4\pi\sqrt{\pi}G\rho a(1-e^2)\left[I+\left(\frac{1+e^2}{2}I-1\right)\frac{\Sigma}{2e^2}\right] \tag{A6}$$

$$C_{20} = \frac{\pi\sqrt{\pi}}{\sqrt{5}}G\rho a\frac{(1-e^2)}{e}\frac{dA}{de}\Sigma, \tag{A7}$$

where,

$$I(e) = \frac{1}{e}\ln\left(\frac{1+e}{1-e}\right) \tag{A8}$$

$$A(e) = 2\frac{1-e^2}{e^2}\left[\frac{1}{2}I-1\right]. \tag{A9}$$

Substituting the coefficients into equations (16) and (17) of Ryden gives the prolate gravitational potential up to quadrupole order.

Table 1: Cluster Observations

| Cluster | Sequence # | Exposure (s) | Background ($10^{-4}$ cts s$^{-1}$ arcmin$^{-2}$) |
|---------|-----------|--------------|-------------------------------------------------|
| A401    | rp800235  | 7465         | 2.07                                            |
|         | wp800182  | 6797         |                                                 |
| A1656   | rp800005  | 22183(20032) | 4.08                                            |
|         | rp800006  | 21893(20136) |                                                 |
|         | rp800009  | 20691(19604) |                                                 |
|         | rp800013  | 22427(20954) |                                                 |
| A2029   | rp800249  | 12550        | 5.50                                            |
| A2199   | wp150083  | 10563        | 3.14                                            |
| A2256   | wp100110  | 17865(14572) | 2.50                                            |
|         | wp800162  | 9108(5380)   |                                                 |
|         | wp800163  | 10803(6690)  |                                                 |
|         | wp800339  | 4978(2437)   |                                                 |
|         | wp800340  | 9430(7119)   |                                                 |
|         | wp800341  | 10480(7469)  |                                                 |

Note. — Time-filtered exposure times are given in parentheses.

Table 2: Pixel Scales

| Cluster | $z$ | Radial Profile (arcsec) | ($h_{80}^{-1}$ kpc) | $D_{max}$ ($h_{80}^{-1}$ kpc) | Ellipticity (arcsec) | ($h_{80}^{-1}$ kpc) | 0.5-2.0 keV Flux ($10^{-12}$ erg cm$^{-2}$ s$^{-1}$) |
|---------|--------|-----|------|------|-----|------|-----------|
| A401    | 0.0748 | 40  | 54.4 | 1251 | 20  | 27.2 | 29.6-29.7 |
| A1656   | 0.0232 | 105 | 44.3 | 487  | 45  | 19.0 | 153       |
| A2029   | 0.0768 | 40  | 55.7 | 1448 | 20  | 27.9 | 40.6-41.0 |
| A2199   | 0.0299 | 90  | 48.9 | 1321 | 30  | 16.3 | 63.8-64.0 |
| A2256   | 0.0581 | 60  | 63.4 | 1458 | 30  | 31.7 | 34.0-34.4 |

Note. — $D_{max}$ is the edge of the outermost bin of the radial profiles. The fluxes are computed in a circle of radius $D_{max}$.

Table 3: $\beta$ Models

| | $a_X$ $(h_{80}^{-1}$ kpc) | | $\beta$ | | | |
| Cluster | bf | 95% | bf | 95% | $\chi^2$ | dof |
|---|---|---|---|---|---|---|
| A401 | 178 | 173-185 | 0.606 | 0.597-0.617 | 27 | 20 |
| A1656 | 267 | 262-273 | 0.785 | 0.771-0.799 | 136 | 8 |
| A2029 | 157 | 141-174 | 0.682 | 0.664-0.700 | 27 | 16 |
| A2199 | 85 | 77-92 | 0.653 | 0.643-0.663 | 56 | 17 |
| A2256 | 397 | 367-427 | 0.908 | 0.870-0.947 | 53 | 9 |

Note. — The 95% confidence limits reflect only statistical uncertainties on two interesting parameters.

Table 4: X-ray Ellipticities and Position Angles (N through E)

| Cluster | $a_{out}$ ($h_{80}^{-1}$ kpc) | $\epsilon_M$ | 95% | $\theta_M$ (deg) | 95% |
|---------|------------|-----|-----|------------|-----|
| A401: | ($a_{in} = 0$) | | | | |
| | 272 | 0.290 | 0.280-0.340 | 29.2 | 18.1-40.9 |
| | 408 | 0.214 | 0.190-0.240 | 33.4 | 29.1-37.8 |
| | 544* | 0.238 | 0.225-0.265 | 38.3 | 35.0-41.7 |
| | 680 | 0.193 | 0.180-0.215 | 39.0 | 35.2-42.7 |
| | 816 | 0.174 | 0.155-0.190 | 37.6 | 33.3-41.0 |
| | 952* | 0.175 | 0.155-0.190 | 33.1 | 28.4-37.9 |
| | 1088 | 0.148 | 0.125-0.165 | 29.2 | 24.3-34.2 |
| A1656: | ($a_{in} = 0$) | | | | |
| | 285 | 0.146 | 0.150-0.155 | 82.7 | 81.4-84.0 |
| | 380 | 0.160 | 0.160-0.165 | 93.2 | 92.7-93.6 |
| | 475* | 0.199 | 0.200-0.205 | 97.7 | 97.2-98.2 |
| A2029: | ($a_{in} = 139$) | | | | |
| | 278 | 0.254 | 0.255-0.290 | 23.3 | 21.2-25.4 |
| | 418 | 0.257 | 0.255-0.285 | 22.1 | 19.7-24.4 |
| | 557* | 0.207 | 0.190-0.225 | 19.6 | 16.4-22.3 |
| | 696 | 0.181 | 0.160-0.205 | 13.8 | 9.99-17.7 |
| | 836 | 0.155 | 0.140-0.180 | 12.5 | 7.88-17.2 |
| | 975* | 0.131 | 0.110-0.160 | 10.9 | 4.23-17.2 |
| | 1114 | 0.116 | 0.090-0.160 | 10.5 | 2.78-18.5 |
| | 1253 | 0.100 | 0.060-0.125 | 2.19 | -8.01-12.2 |
| A2199: | ($a_{in} = 114$) | | | | |
| | 245 | 0.195 | 0.180-0.210 | 42.9 | 39.6-46.1 |
| | 326 | 0.178 | 0.175-0.200 | 40.8 | 37.7-44.0 |
| | 408* | 0.164 | 0.150-0.180 | 36.4 | 33.2-39.5 |
| | 489 | 0.149 | 0.135-0.165 | 36.8 | 33.1-40.7 |
| | 571 | 0.146 | 0.125-0.160 | 34.3 | 30.2-38.1 |
| | 652* | 0.142 | 0.115-0.155 | 34.3 | 29.4-38.8 |
| A2256: | ($a_{in} = 317$) | | | | |
| | 634 | 0.286 | 0.280-0.290 | 122.6 | 122.1-123.1 |
| | 792 | 0.243 | 0.235-0.250 | 123.6 | 122.8-124.3 |
| | 951* | 0.211 | 0.205-0.215 | 123.2 | 121.9-124.4 |
| | 1109 | 0.197 | 0.190-0.205 | 122.1 | 120.9-123.3 |
| | 1268 | 0.185 | 0.170-0.185 | 120.7 | 118.9-122.4 |
| | 1426* | 0.164 | 0.130-0.155 | 119.3 | 116.6-121.4 |

Note. — $a_{in}$ and $a_{out}$ are respectively the inner and outer semi-major axes of the aperture.

Table 5: Shapes of The Total Gravitating Matter

| Cluster | Model | Oblate $\epsilon_{mass}$ | Prolate $\epsilon_{mass}$ |
|---------|-------|--------------------------|---------------------------|
| A401 | C ($\eta = 2$) | 0.52-0.61 | 0.48-0.55 |
| A1656 | C ($\eta = 1$) | 0.40-0.41 | 0.37-0.38 |
| A1656 | C ($\eta = 2$) | 0.42-0.43 | 0.39-0.40 |
| A2029 | H ($n = 3$) | 0.46-0.49 | 0.42-0.44 |
| A2199 | H ($n = 3$) | 0.56-0.64 | 0.49-0.55 |
| A2256 | C ($\eta = 2$) | 0.36-0.38 | 0.34-0.36 |

Note. — These are the 95% confidence uncertainties on $\epsilon_{mass}$ for the isothermal mass models of §3.2.1. The models are C = "cored" power law and H = Hernquist (see §3.1).

Table 6: Effects of Temperature Gradients on Mass Shapes

| $T$ Model | Oblate $\epsilon_{mass}$ | Prolate $\epsilon_{mass}$ |
|-----------|--------------------------|---------------------------|
| $p = -0.27$ | 0.51-0.56 | 0.47-0.50 |
| $p = -0.65$ | $\gtrsim 0.6$ | $\gtrsim 0.6$ |
| $p = +0.07$ | 0.43-0.46 | 0.40-0.42 |
| $p = 0$ | 0.46-0.49 | 0.42-0.44 |

Note. — These are the 95% confidence uncertainties on $\epsilon_{mass}$ for A2029 for different temperature gradients (§3.2.2) where $p$ corresponds to the temperature index in equation [5]. All of the models have Hernquist ($n = 3$) densities (see §3.1) and the $p = 0$ temperature index model is simply the isothermal case given in Table 5.

Table 7: Gas Mass and Composite Mass ($10^{14} M_\odot$)

| | $0.5 h_{80}^{-1}$ Mpc | | | $1.0 h_{80}^{-1}$ Mpc | | | $1.5 h_{80}^{-1}$ Mpc | | |
|---------|-------------|-----------|-----|-------------|-----------|-----|-------------|-----------|------|
| Cluster | $M_{gas}$ | $M_{tot}$ | % | $M_{gas}$ | $M_{tot}$ | % | $M_{gas}$ | $M_{tot}$ | % |
| A401 | 0.276-0.288 | 9.2-10.6 | 3 | 0.863-0.884 | 14.1-25.0 | 3-6 | 1.54-1.64 | 16.0-35.8 | 4-10 |
| A1656 | 0.258-0.259 | 4.4-7.5 | 3-6 | $\cdots$ | $\cdots$ | $\cdots$ | $\cdots$ | $\cdots$ | $\cdots$ |
| A2029 | 0.287-0.308 | 6.0-13.0 | 2-5 | 0.799-0.843 | 11.4-22.5 | 4-7 | 1.38-1.53 | 14.5-26.8 | 5-11 |
| A2199 | 0.144-0.147 | 9.4-12.7 | 1 | 0.387-0.408 | 12.2-17.0 | 2-3 | 0.657-0.711 | 13.2-18.3 | 4-5 |
| A2256 | 0.232-0.256 | 3.4-5.9 | 4-8 | 0.725-0.753 | 9.4-16.8 | 4-8 | 1.16-1.28 | 12.9-22.3 | 5-10 |

Note. — These masses are computed in spheres of radii $(0.5, 1, 1.5) h_{80}^{-1}$ Mpc. The "%" column denotes the percent ratio $M_{gas}/M_{tot}$. The gas mass uncertainties represent the 95% errors of the $\beta$-model parameters and the 90% errors in the temperatures from Edge et al. (1990) as they affect determination of the X-ray flux and plasma emissivity. The composite mass uncertainties represent the 95% statistical uncertainties from the mass models and the 90% errors in the temperatures from Edge et al. (1990).

Table 8: Shapes of The Dark Matter

| Cluster | Model | % $M_{gas}/M_{tot}$ | Oblate $\epsilon_{DM}$ | Prolate $\epsilon_{DM}$ | $r_{gal}$ | $\epsilon_{gal}$ |
|---------|-------|------|------------|------------|-------|-------|
| A401  | C ($\eta = 2$) | 10 | 0.53-0.62 | 0.48-0.56 | 0.625 | 0.46-0.66 |
|       |                | 20 | 0.54-0.63 | 0.49-0.56 |       |           |
| A1656 | C ($\eta = 2$) | 5  | 0.43-0.44 | 0.40-0.41 | 0.313 | 0.51-0.71 |
|       |                | 10 | 0.43-0.44 | 0.41-0.42 | 0.625 | 0.40-0.54 |
| A2029 | H ($n = 3$)    | 10 | 0.47-0.50 | 0.42-0.44 | 0.625 | 0.41-0.61 |
|       |                | 20 | 0.48-0.51 | 0.43-0.45 |       |           |
| A2199 | H ($n = 3$)    | 5  | 0.57-0.65 | 0.50-0.56 | 0.625 | 0.47-0.67 |
|       |                | 10 | 0.58-0.66 | 0.51-0.58 | 1.25  | 0.45-0.63 |
| A2256 | C ($\eta = 2$) | 10 | 0.37-0.39 | 0.34-0.36 | 0.469 | 0.50-0.70 |
|       |                | 20 | 0.39-0.41 | 0.34-0.36 |       |           |

Note. — These are the 95% confidence uncertainties on $\epsilon_{DM}$ for the isothermal mass models of Table 5 where now the models correspond to the dark matter for different ratios of gas mass to total mass. $r_{gal} \equiv \sqrt{ab}$ is the geometric mean radius (in $h_{80}^{-1}$ Mpc) of the aperture used to compute the galaxy isopleth ellipticities, $\epsilon_{gal}$, by Carter & Metcalfe (1980).

---

Fig. 1.— Contour plots of the reduced images for each cluster which have been smoothed with a circular gaussian ($\sigma = 15''$) for display purposes only. The contours are separated by a factor of 2 in intensity and the displayed regions correspond to slightly larger than the region analyzed (i.e. out to radius $D_{max}$). The lowest contour level shown for each cluster corresponds to intensities (in $10^{-4}$ counts cm$^{-2}$ s$^{-2}$) of 1.56 for A401, 40.8 for A1656, 1.77 for A2029, 5.05 for A2199, and 5.00 for A2256.

Fig. 2.— Shown are the background-subtracted radial profiles for each cluster. The top axes are labeled in $h_{80}^{-1}$ kpc and the error bars are placed at the center of each radial bin and the boxes indicate bins omitted from analysis.

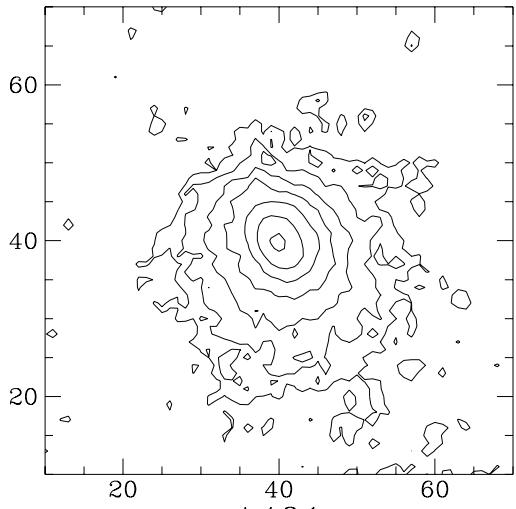

A401

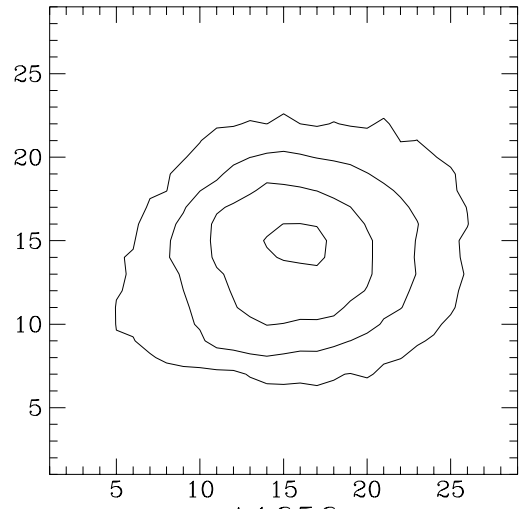

A1656

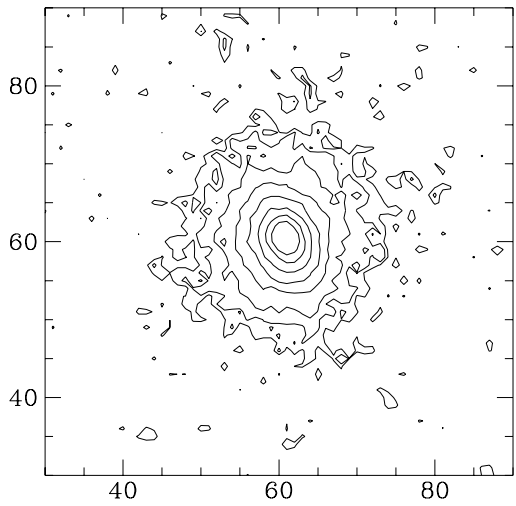

A2029

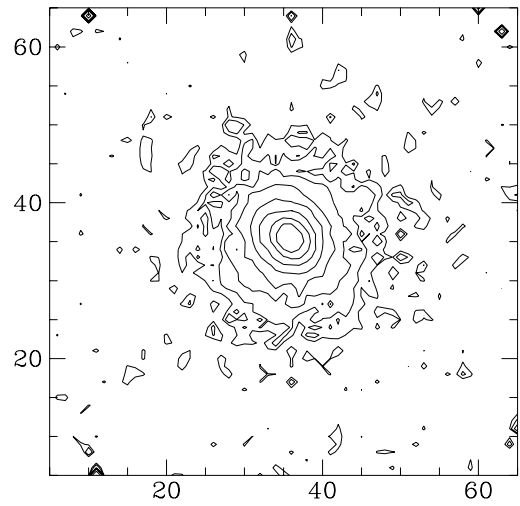

A2199

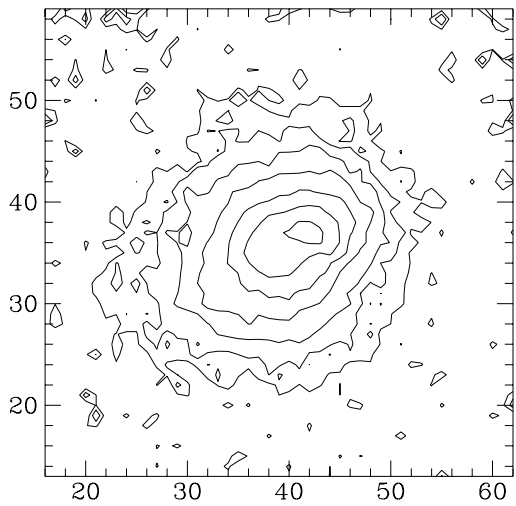

A2256

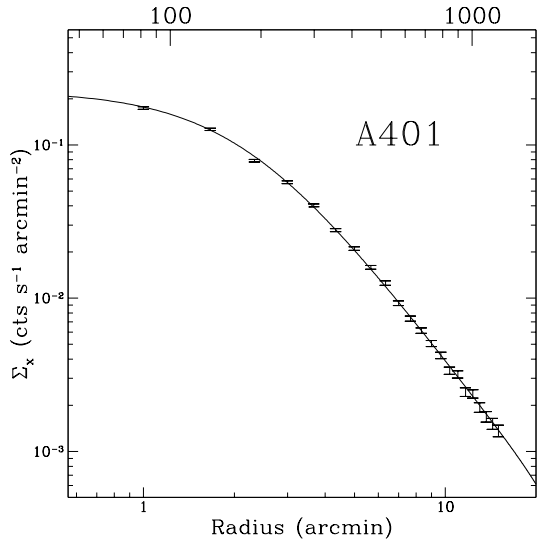

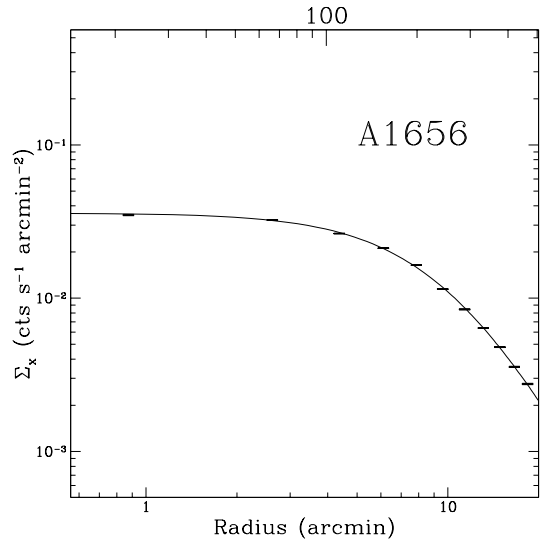

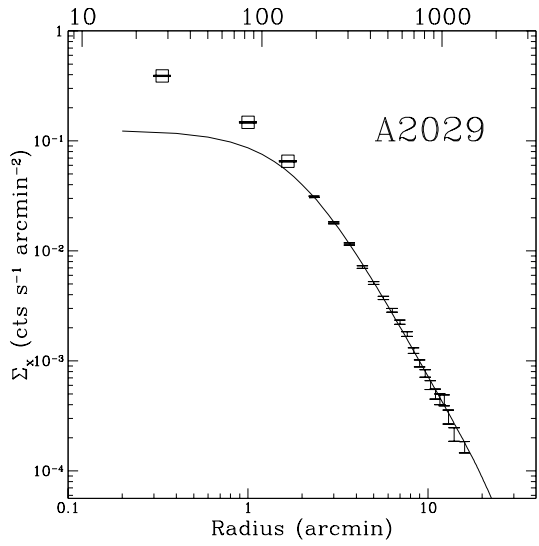

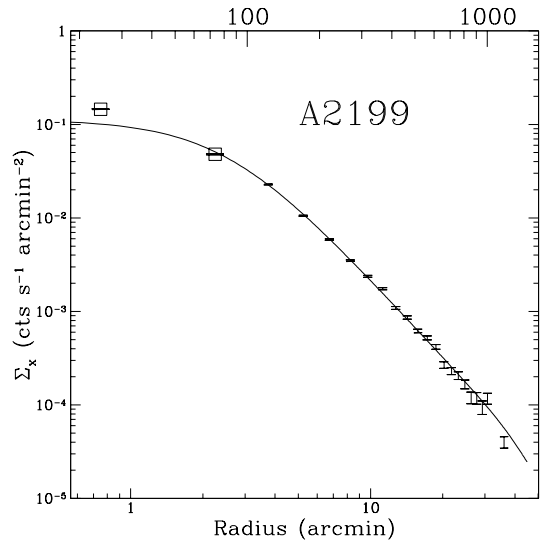

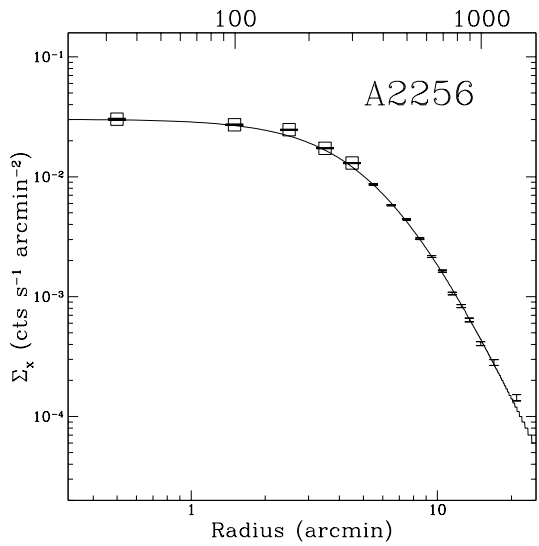